\documentclass[twocolumn]{aastex63}
\usepackage{natbib}
\usepackage{subfigure}
\usepackage{amsmath}
\newcommand{\msun}{{M_{\odot}}}
\newcommand{\mstar}{{M_{\ast}}}
\newcommand{\ser}{S\'ersic}

\received{April 13, 2022}
\revised{June 1, 2022}
\accepted{June 3, 2022}
\published{June 17, 2022}

\shorttitle{No Half-Mass Size Evolution of the Milky-Way Type Galaxies}
\shortauthors{Hasheminia et al.}

\begin{document}

\title{No Evolution in the Half-mass Radius of Milky Way-type Galaxies over the Last 10 Gyr}

\correspondingauthor{Moein Mosleh}
\email{moein.mosleh@shirazu.ac.ir}

\author[0000-0003-3428-6441]{Maryam Hasheminia}
\affiliation{Biruni Observatory, School of Science, Shiraz University, Shiraz 71946-84795, Iran}
\affiliation{Department of Physics, School of Science, Shiraz University, Shiraz 71946-84795, Iran}
\affiliation{Department of Physics, Institute for Advanced Studies in Basic Sciences (IASBS), P.O. Box 11365-9161, Zanjan, Iran}
\author[0000-0002-4111-2266]{Moein Mosleh}
\affiliation{Biruni Observatory, School of Science, Shiraz University, Shiraz 71946-84795, Iran}
\affiliation{Department of Physics, School of Science, Shiraz University, Shiraz 71946-84795, Iran}
\author[0000-0002-8224-4505]{Sandro Tacchella}
\affiliation{Kavli Institute for Cosmology, University of Cambridge, Madingley Road, Cambridge, CB3 0HA, UK}
\affiliation{Cavendish Laboratory, University of Cambridge, 19 JJ Thomson Avenue, Cambridge, CB3 0HE, UK}
\author[0000-0003-3449-2288]{S. Zahra Hosseini-ShahiSavandi}
\affiliation{Biruni Observatory, School of Science, Shiraz University, Shiraz 71946-84795, Iran}
\author[0000-0002-8435-9402]{Minjung Park}
\affiliation{Center for Astrophysics $\mid$ Harvard \& Smithsonian, 60 Garden Street, Cambridge, MA 02138, USA}
\author[0000-0003-3997-5705]{Rohan P. Naidu}
\affiliation{Center for Astrophysics $\mid$ Harvard \& Smithsonian, 60 Garden Street, Cambridge, MA 02138, USA}

\begin{abstract}
The Milky Way (MW) galaxy is in focus, thanks to new observational data. Here we shed new light on the MW's past by studying the structural evolution of MW progenitors, which we identify from extragalactic surveys. Specifically, we constrain the stellar-mass growth history (SMGH) of the MW with two methods: ($i$) direct measurement of the MW's star formation history, and ($ii$) assuming the MW is a typical star-forming galaxy that remains on the star-forming main sequence. We select MW progenitors based on these two SMGHs at $z=0.2-2.0$ from the CANDELS/3D-HST data. We estimate the structural parameters (including half-mass radius $r_{50}$ and \ser index) from the stellar-mass profiles. Our key finding is that the progenitors of the MW galaxy grow self-similarly on spatially resolved scales with roughly a constant half-mass radius ($\sim2-3$ kpc) over the past 10 Gyr, while their stellar masses increase by about 1 dex, implying little-to-no inside-out growth. We discover that the radius containing $20\%$ of the stellar mass ($r_{20}$) decreases by $60\%$ between redshifts of $z=2.0$ and $z=0.7$, while the central stellar-mass density ($\Sigma_1$) increases by a factor of 1.3 dex over the same time and the \ser index changes as $n\propto(1+z)^{-1.41\pm0.19}$. This is consistent with an early ($z>1$) formation of a thick disk, followed by the formation of a bar that led to an increase in the mass in the core. The formation and evolution of the thin disk had only little impact on the overall half-mass size. We also show that the constant-size evolution of the MW progenitors challenges semiempirical approaches and numerical simulations.
\end{abstract}


 

\section{Introduction}
\label{introduction}

Understanding the assembly history of the Milky Way (MW) galaxy is of great interest because we live within it and because the MW—as a typical disk galaxy—provides a unique probe of galaxy evolution. Thanks to invaluable data from Gaia \citep{GaiaCollaboration2018} and other surveys \citep[e.g.,][]{Conroy2019}, recent studies allowed us to put better constraints on the stellar-mass growth history (SMGH) of the MW and produced a picture in which most of the MW stellar mass is assembled in situ over the last 8--11 Gyr \citep[e.g.,][]{Belokurov2018, Helmi2018, Bonaca2020, Kruijssen2020}. In this work, we connect these recent insights to extragalactic studies to shed new light on the structural evolution of the MW and disk galaxies in general.

Observationally, it remains challenging to understand how individual galaxies assemble and evolve about scaling relations \citep[e.g.,][]{Abramson2016, Tacchella2016}. In particular, regarding the structural evolution of galaxies, recent observational studies compared the structural parameters of galaxies at a fixed stellar mass over time \citep[e.g.,][]{vanderWel2014, Mowla2019}. However, the evolution of galaxies should be traced by linking progenitors and descendants through cosmic time \citep[e.g.,][]{Carollo2013}.

Several different approaches have been used to select progenitors. One such approach is the constant cumulative number density method, which assumes that the galaxies have a fixed comoving number density and rank order at all epochs, thereby ignoring mergers and variations in star formation \citep{VanDokkum2013}. Variations in the SMGH can partially be addressed by abundance matching, which leads to an evolving cumulative number density to compare galaxies across cosmic time \citep[][]{Behroozi2013, Moster2013, Papovich2015, Torrey2015}. In this method, observed galaxies are matched to dark matter halos for which the number density evolution can be tracked through simulations. In a third method, as in situ star formation dominates the mass assembly of star-forming galaxies (SFGs), one can trace the SMGH of SFGs from the observed star formation history (SFH) \citep[][]{Renzini2009, Leitner2011, Patel2013}. Because we focus on the MW for which we have good knowledge of the SMGH over the last $\sim10$ Gyr and evidence for mostly in situ star formation \citep[e.g.,][]{Kruijssen2019, Bonaca2020}, we adopt this last method in this work.

Concerning the morphology of galaxies, several recent studies have shown that considering the effect of mass-to-light ($M_{\ast}/L$) gradients and hence analyzing the stellar-mass profiles instead of their light profiles is essential to better constrain the structural evolution of galaxies \citep{Tacchella2015b, Mosleh2017, Mosleh2020, Suess2019a, Suess2019b}. Moreover, different size definitions can be used to depict changes in subcomponents of galaxies \citep{Cibinel2013, Miller2019}. Therefore, in this study, we carefully select MW progenitors and use the mass-based structural parameters to revisit the buildup of the MW stellar-mass profile (Section~\ref{samples}). We find negligible evolution in the half-mass size of MW-type galaxies over the past 10 Gyr (Section~\ref{results}), implying little-to-no inside-out growth of such disk galaxies (Section~\ref{discussion}). We show that this is largely independent of the assumed SMGH as the half-mass size versus stellar-mass relation is flat and nearly redshift invariant.

\begin{figure*}
	\includegraphics[width=\textwidth]{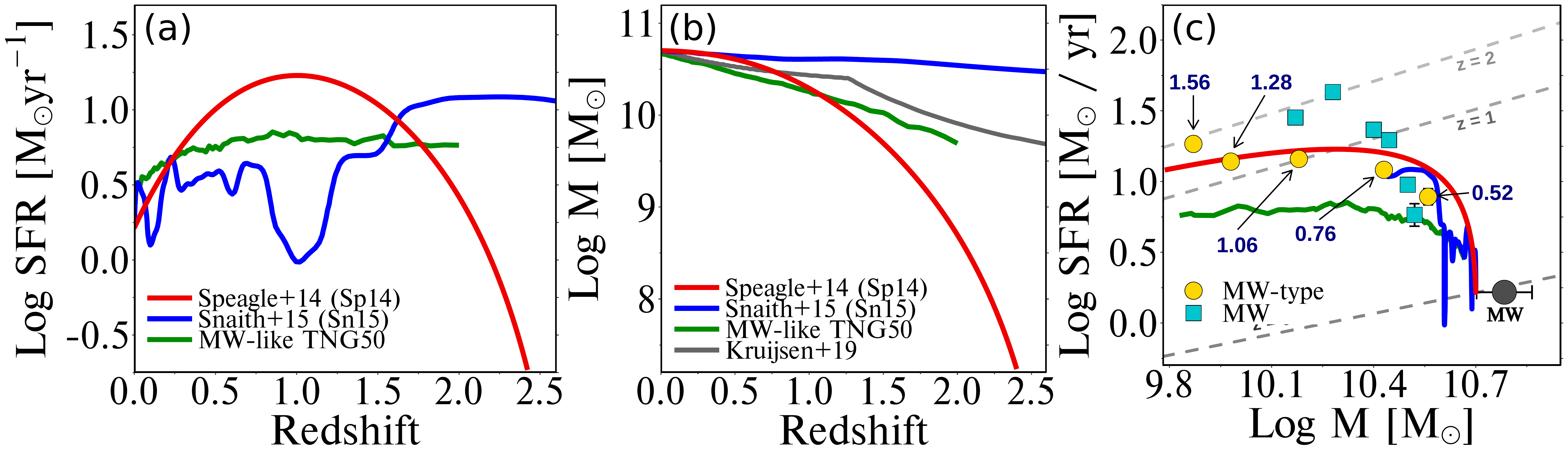}
	\caption{(\textit{a}) SFH of ($i$) MW-type galaxies derived by applying the MSI method to the $\mathrm{SFR}-\mstar$ relation of Sp14 (red line); ($ii$) the MW from Sn15 (blue line); and ($iii$) MW-like galaxies in the IllustrisTNG (TNG50) simulation (green line). (\textit{b}) Corresponding SMGHs with a final stellar mass of $M_{\ast,0} = 10^{10.7} \msun$. The SMGH of the MW of Sn15 is in good agreement with \cite{Kruijssen2019}, which is shown in gray. (\textit{c}) Tracks of the MW-type galaxies (red line), the MW (blue line), and the TNG MW-like galaxies (green line) in the $\mathrm{SFR}-\mstar$ plane. The yellow circles and cyan squares are the median values of the M$_{\star}$-selected MW-type and MW progenitors at the different redshifts (indicated with the labels). The dashed gray lines indicate the $\mathrm{SFR}-\mstar$ relation of Sp14. The large gray circle is the M$_{\star}$ and SFR of the MW estimated by \cite{Licquia2015}.}
	\label{fig1}
\end{figure*}
\section{Data and Sample Selection}
\label{samples}

\subsection{Observational Data}
\label{data}

We aim at selecting star-forming progenitors of MW-type galaxies with a total stellar mass of $\approx5\times10^{10}~\msun$ in the present-day universe \citep{McMillan2017}. We use the CANDELS/3D-HST catalogs \citep{Skelton2014} to identify those progenitors at earlier cosmic times. We cross-matched this sample with the \cite{Mosleh2020} catalog, which provides stellar-mass profiles and mass-based structural parameters for 5557 galaxies. These have been obtained by producing the stellar-mass maps from the Hubble Space Telescope (HST) imaging data and applying the pixel-by-pixel spectral energy distribution fitting technique. The reliability of the technique has been examined in \cite{Mosleh2020} using simulated galaxies, and it has been shown that the measured parameters are robust and the sample is $\ga90\%$ complete for galaxies with $\log(\mstar/\msun)\geqslant9.8$ at $0.2\leqslant z\leqslant2$.

\subsection{Stellar Mass Growth History (SMGH)}
\label{MSI}

To select the MW progenitors, we derive the SMGHs in two different ways: ($i$) using the main-sequence integration (MSI) method developed by \cite{Renzini2009, Leitner2011} to follow SFGs on the star-forming main sequence (referred to as MW-type SMGH) and ($ii$) adopting the inferred SFH of MW studies (referred to as MW SMGH).

In the first approach (MW-type SMGH), the basic assumption is that present-day SFGs (including the MW) have assembled most of their stellar masses steadily and smoothly by in situ star formation. Thus, the star formation rate (SFR) of MW-type progenitors must fall within the scatter of the observed star-forming main sequence ($\mathrm{SFR}-\mstar$ relation). The SMGH can then be iteratively calculated by obtaining the SFR from the evolving $\mathrm{SFR}-\mstar$ relation and taking into account stellar-mass loss. In this work, we apply the $\mathrm{SFR}-\mstar$ relation of \citep[hereafter Sp14]{Speagle2014} and that the final ($z=0$) stellar mass of the MW is $\mstar=10^{10.7}~\msun$ \citep{McMillan2017}. 

In the second approach (MW SMGH), we employ the SFH estimates of MW studies to determine its SMGH directly. In this work, we use the SFH of \citet[hereafter Sn15]{Snaith2015} which is obtained from fitting the solar vicinity $age-[\alpha/Fe]$ relation to a chemical evolution model. The SMGH derived by this SFH is in good agreement with the MW mass assembly history presented by \cite{Kruijssen2019}.

To compare our observational results with numerical simulation, we select MW-like galaxies from the IllustrisTNG (TNG50) simulation \citep{Nelson2019, Pillepich2019} in the following way at $z=0$ \citep[see][for the definition of those quantities]{Park2021}:

\begin{itemize}
    \item Stellar mass ($<30$ kpc): $\mstar=10^{10.5}-10^{10.9}\msun$;
    \item Disk-to-total ratio: $D/T>0.5$;
    \item $\mathrm{SFR}/(M_{\odot}\/\mathrm{yr}^{-1})=[0.1, 10.0]$.
\end{itemize}

Figure~\ref{fig1} shows in panel (\textit{a}) and panel (\textit{b}) the SFH and SMGH of MW-type galaxies inferred from the MSI method with the $\mathrm{SFR}-\mstar$ relation of Sp14 (red lines) of the MW galaxy from Sn15 (blue lines) and of MW-like galaxies from the TNG50 simulation (green lines). The SMGH from the MW indicates an earlier formation (consistent with \citealt{Kruijssen2019}) than the one from the MSI method. Figure~\ref{fig1}(c) shows the evolution of the SFR and $\mstar$. The large gray circle is the MW mass and SFR estimated by \cite{Licquia2015}, highlighting that our SMGHs are consistent with this estimate.

\subsection{Selection of Progenitors}
\label{selection}

We select the MW progenitors at a given redshift to be all SFGs with a stellar mass consistent within $\pm0.1$ dex of the MW and MW-type SMGHs inferred above. SFGs are defined by their location in the rest-frame UVJ color diagram with the boundaries given in \cite{Mosleh2017}. In order to reduce the uncertainties, we exclude galaxies with unreliable structural parameters \citep[see][for more details]{Mosleh2020}. A total of 1250 and 721 objects are identified as possible MW-type and MW progenitors up to $z\sim1.7$ and $z\sim2$, respectively.
In Figure~\ref{fig1}(\textit{c}), the yellow circles and cyan squares show the median of the integrated UV+IR SFRs from \citet{Whitaker2014} for the MW-type and MW progenitors, respectively. The yellow circles are different from the solid red line because the \textit{UVJ}-selected SFGs do not exactly reproduce the $\mathrm{SFR}-\mstar$ relation of Sp14. The cyan squares differ from the blue line because the \textit{UVJ}-selected SFGs have typically higher SFRs than the MW at fixed stellar mass and redshift.

\section{Results}
\label{results}
\begin{figure*}
\includegraphics[width=\textwidth]{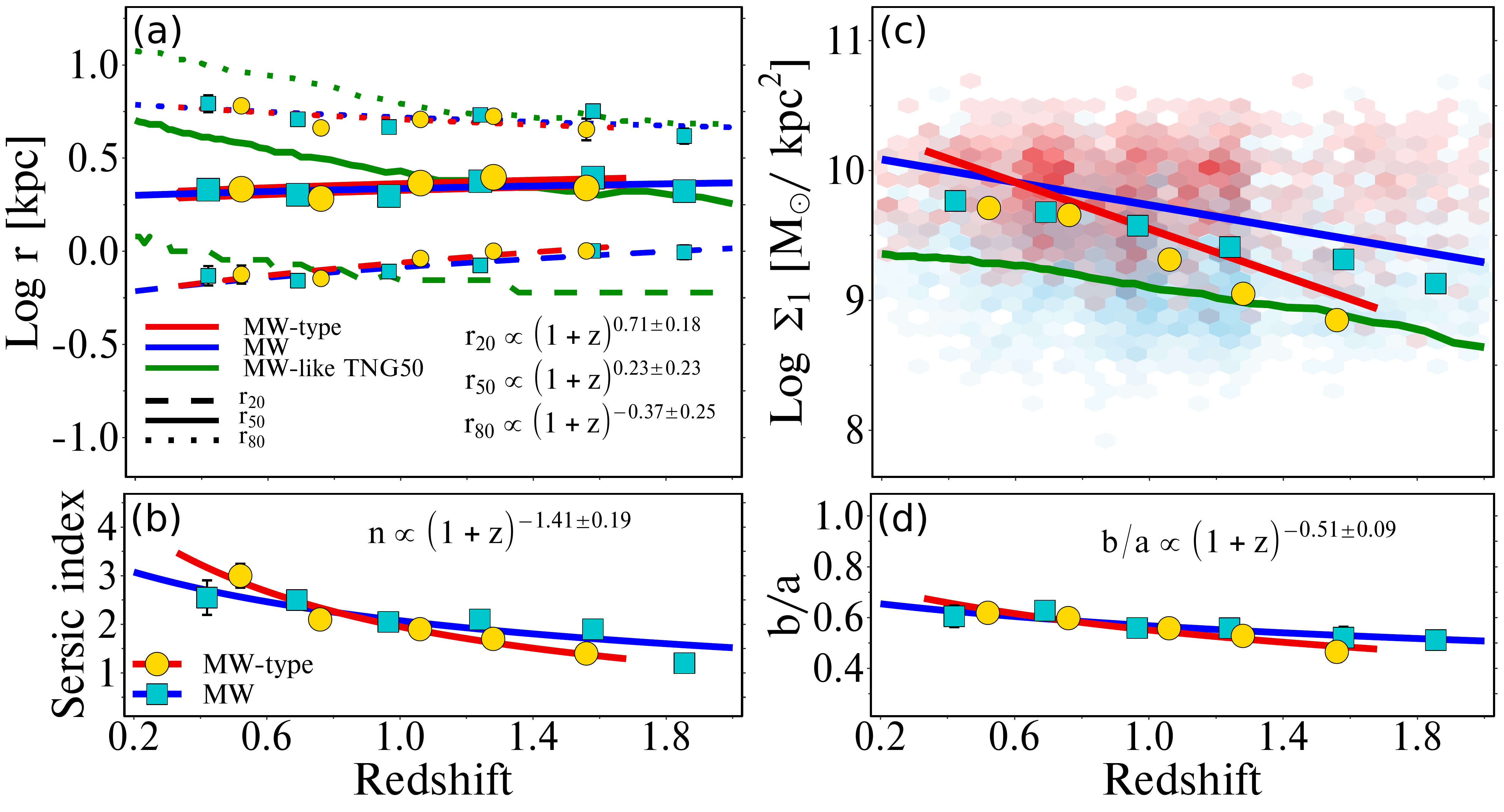}
\caption{Evolution of the size (panel (\textit{a})), \ser index (panel (\textit{b})), central stellar-mass density $\Sigma_1$ (panel (\textit{c})), and axis ratio (panel (\textit{d})) with redshift for MW progenitors. The yellow circles and cyan squares are for the MW-type and MW progenitors, respectively, and the solid red and blue lines are the corresponding best fits. The green solid lines show the evolution of the TNG50 MW-like galaxies. In panel (\textit{a}), dashed (small symbols), solid (large symbols), and dotted (small symbols) lines indicate the median evolution of the radii that enclose 20\% ($r_{20}$), 50\% ($r_{50}$), and 80\% ($r_{80}$) of the stellar mass, respectively. In panel (\textit{c}), the red and blue hex-histograms show the $\Sigma_1$ distribution of quiescent galaxies and SFGs of a stellar-mass-complete sample. A considerable decrease in $r_{20}$ and an increase in the \ser index and $\Sigma_1$ are observed with cosmic time, whereas $r_{50}$ and $r_{80}$ remain approximately constant. In contrast, the TNG50 MW-like galaxies show a significant increase in $r_{20}$, $r_{50}$ and $r_{80}$ with cosmic time.}
\label{fig2}
\end{figure*}

\subsection{Size Evolution}

Figure~\ref{fig2}(\textit{a}) shows the size growth of the MW-type progenitors (yellow symbols) and the MW progenitors (cyan symbols). We focus on the stellar-mass-based radii that enclose 20\% ($r_{20}$), 50\% ($r_{50}$), and 80\% ($r_{80}$) of the total stellar mass. We fit a power law of the form $r\propto(1+z)^\gamma$ to the median values to quantify the rate of size evolution.

Focusing on the evolution of MW-type galaxies (red lines and yellow circles), the half-mass sizes ($r_{50}$) of the progenitors remain almost constant within the redshift range of this study. The rate of $r_{80}$ size growth is also shallow ($\gamma=-0.37\pm0.25$). By contrast, the $r_{20}$ size decreases with a rate of $\gamma=0.71\pm0.18$. Similar results are obtained if we focus on the MW progenitors (blue lines and cyan squares). In contrast, the sizes of MW-like galaxies from TNG50 (green lines) increase significantly with cosmic time, particularly below $z\sim1$.

The slow rate of the half-mass size evolution for SFGs has already been alluded to by previous studies that performed the analysis at fixed masses \citep[e.g.,][]{Mosleh2020, Suess2019a}. However, tracing progenitors makes this even more apparent. This might be caused by the simultaneous buildup of stellar mass in the core and the outskirts at large radii over time. This self-similar growth of the stellar-mass profile is consistent with a flat radial specific SFR profile, as observed at intermediate redshifts \citep[e.g.,][]{Tacchella2015a, Tacchella2018, Nelson2016, Morselli2019}.

\begin{figure*}
\includegraphics[width=\textwidth]{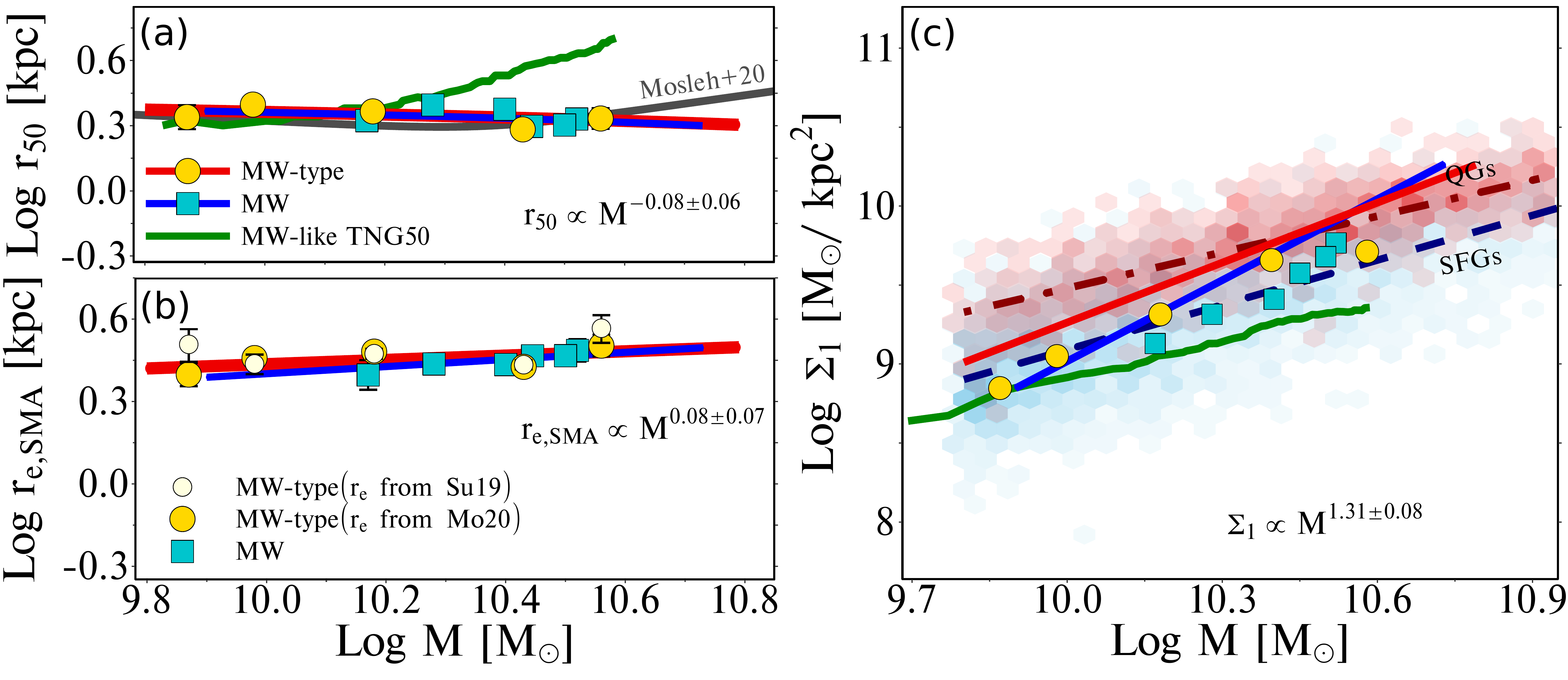}
\caption{Evolution of the median circularized half-mass size $r_{50}$ (panel (\textit{a})), half-mass semimajor axis size $r_{\rm e,SMA}$ (panel (\textit{b})), and central stellar-mass density $\Sigma_1$ (panel (\textit{c})) as a function of mass. Lines and symbols are the same as in \ref{fig2}. The gray solid line in panel (\textit{a}) indicates the size--mass relation of \cite{Mosleh2020} for SFGs at $0.3<z<0.7$, while the dashed navy and dashed--dotted dark-red lines in panel (\textit{c}) mark the $\Sigma_1-\mstar$ relation for SFGs and QGs. The small beige circles in panel (\textit{b}) show the size--mass evolution when adopting the size measurements of \citet{Suess2019a}. MW-type progenitor galaxies exhibit a steeper evolution in the $\Sigma_1-\mstar$ plane than SFGs, i.e., they have a faster central mass buildup than SFGs, consistent with the $r_{20}$ evolution.} 
\label{fig3}
\end{figure*}

This pattern of self-similar growth can also be seen directly in Figure~\ref{fig3}(\textit{a}), where we plot the $r_{50}-\mstar$ plane. The MW and MW-type progenitors increase their stellar mass by nearly 1 dex, while their $r_{50}$ size remains constant ($r\propto M^\beta$ with $\beta=-0.08\pm0.06$). Their evolution tracks well with the $r_{50}-\mstar$ relation of\citet{Mosleh2020} for SFGs at $z=0.3-0.7$, which is not too surprising because this relation does not evolve strongly with cosmic time.

To check whether our results depend on the exact size definition (above, we have used circularized half-mass sizes) and methodology, we plot in Figure~\ref{fig3}(\textit{b}) the half-mass radius of the semimajor axis $r_{\rm e,SMA}$ from this work and \cite{Suess2019a}. Adopting $r_{\rm e,SMA}$ instead of $r_{50}$ leads to a weak trend in the opposite direction, i.e., $r_{\rm e,SMA}$ weakly increases with mass and time ($\beta=0.08\pm0.07$), but it is still consistent with being constant. This is expected because the axis ratio $b/a$ increases with cosmic time for MW and MW-type progenitors (Figure~\ref{fig2}(\textit{d})). Adopting the size measurements of \cite{Suess2019a} instead of the ones of \citet{Mosleh2020} leads to negligible changes. We conclude that our key result of negligible size growth of MW and MW-type progenitors is robust.

\subsection{Growth of Central Densities}

The buildup of the central density can be directly linked to an increase of the \ser index ($n$) and the stellar-mass surface density within 1 kpc ($\Sigma_1$). Panels (\textit{b}) and (\textit{c}) of Figure~\ref{fig2} show the redshift evolution of the \ser index $n$ and $\Sigma_1$ for the MW-type (yellow circles) and MW progenitors (cyan squares).

We find that both the MW and MW-type progenitors increased $n$ and $\Sigma_1$ with cosmic time, with the trend being shallower for the MW than the MW-type progenitors. For the \ser index $n$, we find $n\propto(1+z)^{-1.41\pm 0.19}$ and $n\propto(1+z)^{-0.77\pm 0.21}$ for the MW-type and MW progenitors, respectively. Furthermore, MW-type progenitors increase $\Sigma_1$ by nearly one order of magnitude from $z\approx1.7$ to $z\approx0.7$, while a shallower evolution is found at lower redshifts, independent of the parametric or nonparametric $\Sigma_1$ estimation. This is consistent with the increase in $n$ and the decrease of $r_{20}$, pointing to a significant buildup of mass in the core, reaching values of $\Sigma_1$ that are consistent with quiescent galaxies (QGs). Interestingly, $n$ reaches a value of only $n\approx2$ by $z\approx0.7$, compatible with a pseudo-like bulge or a thick disk rather than a classical bulge component. We caution that the results are based on a single \ser model. The MW-like progenitors of the TNG50 simulation (green line in Figure~\ref{fig2}(\textit{c})) show a shallower evolution in $\Sigma_1-z$ than one of the MW-type progenitors but are roughly consistent with the MW progenitors.

The track of progenitors in the $\Sigma_1$--mass plane is illustrated in Figure~\ref{fig3}(\textit{c}). Compared with the general relation of SFGs (dashed navy line), the MW-type and MW progenitors trace steeper relations as indicated by the red ($\Sigma_1\propto M^{1.31\pm0.08}$) and blue ($\Sigma_1\propto M^{1.71\pm0.19}$) lines, respectively. For comparison, the SFGs and QGs have slopes of $0.94\pm0.01$ and $0.76\pm0.01$, respectively, in agreement with \cite{Barro2017}.

\begin{figure*}
\includegraphics[width=\textwidth]{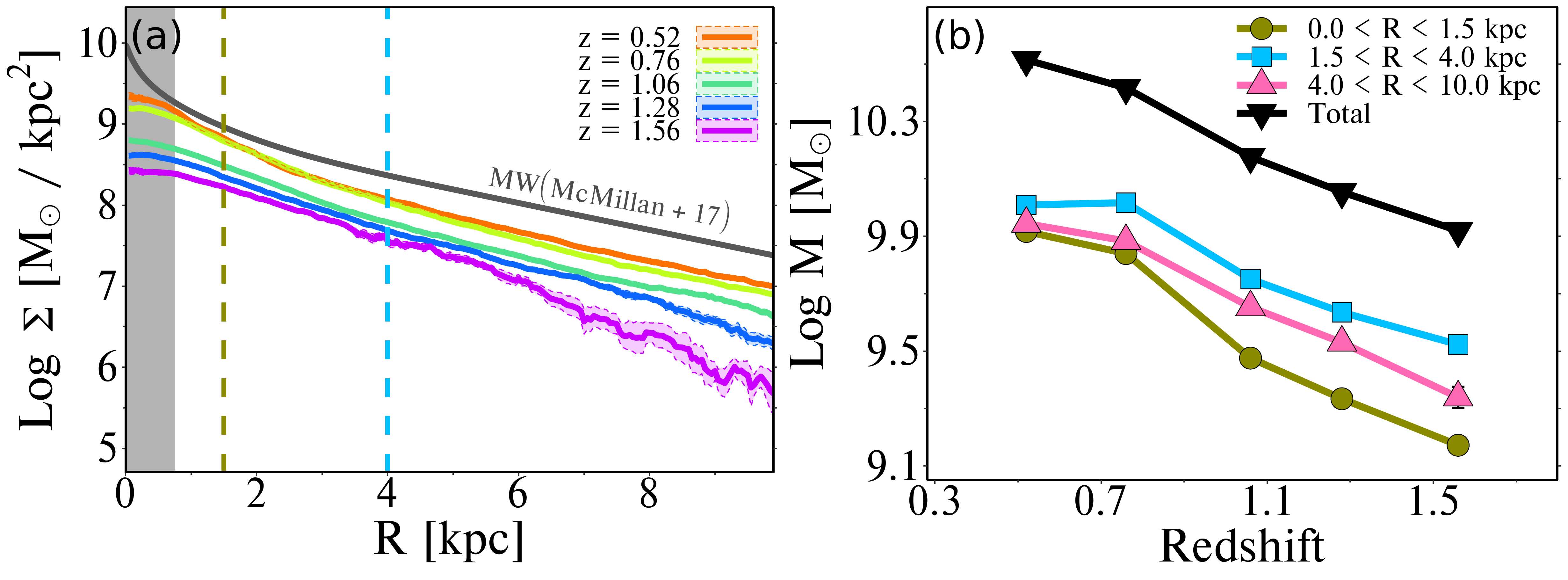}
\caption{(\textit{a}) Comparing the median stellar-mass density profiles of MW-type progenitors, along with the profiles of the MW from the best-fit parameters estimated by \cite{McMillan2017}. (\textit{b}) The total stellar-mass growth within different apertures. The stellar mass in the core has more rapidly increased than that in the other regions from $z=1.1$ to $z=0.7$, consistent with the decrease in $r_{20}$ and increase in $\Sigma_1$ over the same epoch (Figure~\ref{fig2}).}
\label{fig4}
\end{figure*}

Finally, for the MW-type progenitors, we tested whether the profile change is self-similar at all redshifts by comparing the shape of their median mass profiles (Figure~\ref{fig4}). As a comparison, we used the MW mass profile from \cite{McMillan2017}, noting that they assumed a spatially constant $\mstar/L$. Figure~\ref{fig4}(\textit{a}) shows that the shape and slope of the profiles mostly remain the same with time, though there is some change in the central regions ($\lesssim1.5$ kpc) at later times. The growth of stellar mass within different apertures is shown in panel (\textit{b}). The MW-type progenitors grow largely self-similarly, though there is a phase of accelerated core growth between $z=1.1$ and $z=0.7$. At late times ($z<0.7$), there seems to be more growth beyond 10 kpc than within.

\begin{figure*}
\includegraphics[width=\textwidth]{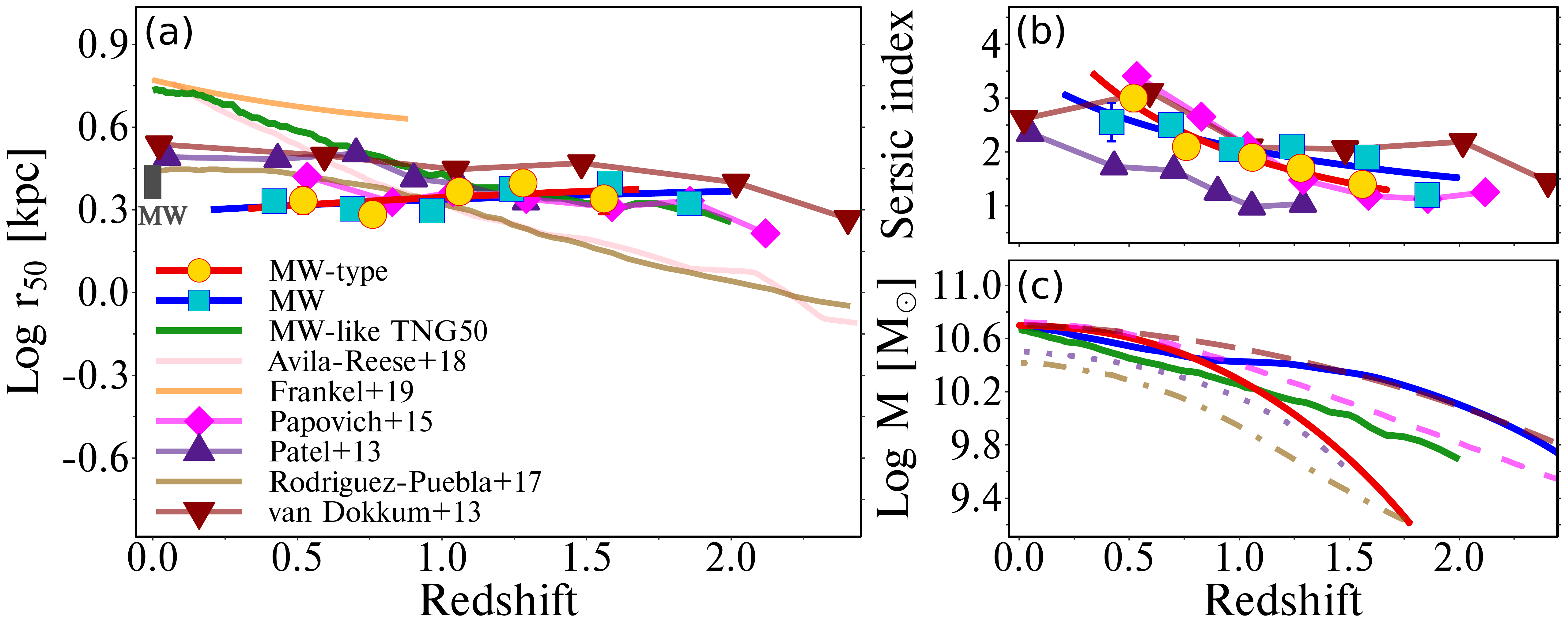}
\caption{Comparison of the redshift evolution of $r_{50}$ (panel (\textit{a})), \ser index (panel (\textit{b})), and stellar mass (panel (\textit{c})) of MW progenitors with recent observational and theoretical studies. Lines and symbols follow Figures\ref{fig1} and \ref{fig2}. The gray rectangle represents the MW's $r_{50}$ derived from the \cite{McMillan2017} profile and corrected with the typical $r_{\rm mass}/r_{\rm light}$ from \cite{Mosleh2017}. Overall, a consistent picture for the MW galaxy arises: While the stellar mass has increased by $>1$ order of magnitude over the past 10 Gyr, the half-mass size has roughly stayed constant. Over the same time span, the \ser index increased from disk like (close to 1) to 2--4 at lower redshift, consistent with the formation of a thick disk or bulge.}
\label{fig5}
\end{figure*}
\section{Discussion \& Conclusions}
\label{discussion}

Our key result, which is summarized and compared with other observational \citep{Patel2013, VanDokkum2013, Papovich2015}, theoretical \citep{Rodriguez2017, Frankel2019}, and simulated \citep{Avila-Reese2018} studies in Figure~\ref{fig5}, is that the progenitors of the MW galaxy grow self-similarly and have roughly a constant half-mass radius over the past 10 Gyr, while their stellar mass increases by about 1 dex. This holds while adopting two rather different SMGHs: one obtained via the MSI method (MW type; red lines) and one measured directly from the MW (MW; blue lines).

\cite{Patel2013}, also using the MSI method, find a weak increase in $r_{50}$ at the late cosmic time, while the increase in their \ser index is weaker and with a lower normalization than our estimate. Adopting the $\mathrm{SFR}-\mstar$ relation of their work, we find our results still hold. We therefore attribute these differences to the increasing importance of the $\mstar/L$ gradient at late times \citep[e.g.,][]{Mosleh2017, Suess2019b}. \cite{VanDokkum2013} and \cite{Papovich2015} adopt different selection techniques to identify MW progenitors (constant comoving number density and abundance matching, and they both include QGs), which lead to shallower SMGHs than those estimated by the MSI method (Figure~\ref{fig5}(\textit{c})). Their inferred size growth shows a slow rate of evolution, similar to our study, despite using half-light radii. They also find that bulges have already been built by $z\sim0.5$ (Figure~\ref{fig5}(\textit{b})).

\cite{Rodriguez2017} adopt a semiempirical approach for connecting galaxies to their host halos. The SMGH for the final halo mass of $10^{12}\msun$ of the MW shows a similar trend to this work (Figure~\ref{fig5}(\textit{c})). However, the inferred size evolution, which assumes that the sizes scale as $\propto H(z)^{-0.5}$, predicts a steeper trend, particularly at $z>1$. Moreover, \cite{Frankel2019} estimate the half-mass size evolution of the MWs low-$\alpha$ disk ($R=6-13$ kpc) by fitting a global model to the ages, metallicities, and radii of APOGEE red clump stars and find a size growth of $r_{1/2}\propto(1+z)^{-0.49}$, which is in line with our $r_{80}$ size growth, but significantly larger than our $r_{50}$ estimates. We also find that the size evolution of MW-like galaxies in the TNG50 simulation (green line in Figure~\ref{fig5}) is consistent with our results at early times but shows a strong increase in more recent times ($z<0.7$), which is contrary to our findings. This rapid increase in size is similar to that driven by \cite{Avila-Reese2018}, who simulated the evolution of eight zoomed-in hydrodynamical MW-sized galaxies. As seen in Figure~\ref{fig5}(\textit{a}), $r_{50}$ grows relatively faster than in any observational study.

A concern of our analysis is the assumption that mergers play no significant role in the buildup of the stellar mass of the MW. \cite{Kruijssen2020} and \cite{Naidu2021} estimate that the most massive mergers of the MW contributed a total stellar mass of $\log(M/\msun)\approx10^9$, suggesting that the MW formed mainly by in situ star formation. Therefore, we conclude that mergers did not significantly affect the mass assembly of the MW progenitor, at least for the redshift range of this study.

Our analysis also has important consequences regarding the angular momentum history of galactic disks (such as the MW). The angular momentum of a galaxy is
\begin{equation}
    J_{\star}\propto M_{\star}r_{50}v_{\rm rot},
\end{equation}
where $v_{\rm rot}$ is the rotational velocity, which increased from $z\approx2$ to today for MW-type galaxies by a factor of $\sim1.5$ \citep{Simons2017}. As discussed above, the stellar mass of the MW increased by a factor of 5--10 from $z\approx2$ to today, while the half-mass size remains constant. Putting this all together, we find that the angular moment of MW-like disks increased over the past 10 Gyr by a factor of $J(z=0)/J(z=2)\approx8-15$. This estimate is consistent with canonical dark matter theory \citep{Peebles1969, Danovich2015, DuttonBosch2012} and not as extreme as recently claimed by \cite{PengRenzini2020} and \cite{Renzini2020}, who have used observational scaling relations and inferred an increase by a factor of $\sim20-50$. In the future, direct estimates of the angular momentum history of the MW will provide important constraints on the vorticity of the accreted gas and accretion history of satellite galaxies.

From the MW perspective, these results seem to be in accordance with the early ($z>1$) formation of the thick disk, followed by a decrease in the SFR and a formation of a bar \citep{Haywood2018, Bovy2019}. This bar redistributed the stellar mass of the thick disk, thereby increasing the mass surface density in the core. This is consistent with our inferred increase of the Sérsic index and central stellar-mass density at $0.7<z<1.5$. Moreover, the mild growth of our half-mass size, in particular of $r_{e,SMA}$, is compatible with the slow thin-disk formation over the last $7-8$ Gyr \citep{Conroy2022}.

As shown here, connecting detailed studies of the MW galaxy with extragalactic studies of MW-like galaxies is fruitful. Upcoming telescopes (including the James Webb and the Roman Space Telescopes) will probe MW-like galaxies to higher redshifts and with better spatial resolution, allowing us to constrain the cosmic evolution of the thickness of disks and the occurrence of bars.

We thank the anonymous referee for the suggestions that helped improve the manuscript. We are also grateful to Alvio Renzini for insightful comments.
\bibliographystyle{apj}


\end{document}